\documentclass[a4paper,11pt]{article}
\pdfoutput=1

\usepackage{lineno}

\usepackage{jcappub} 

\usepackage[T1]{fontenc}

\usepackage{graphicx}
\usepackage{subfig}
\usepackage{hyperref}
\usepackage{amsmath}

\usepackage[utf8]{inputenc}
\newcommand{\lnls}{log$N$-log$S$}
\title{The promise of deep-stacking for neutrino astronomy} %or Blueprint for  identifying the origin of extra-galactic high-energy  neutrinos and cosmic rays}
\author[a,b]{Marek Kowalski} 
\author[a]{Markus Ackermann} 
\author[c]{Imre Bartos}
\affiliation[a]{Deutsches Elektronen Synchrotron DESY, Platanenallee 6, 15738 Zeuthen, Germany}
\affiliation[b]{Institut für Physik, Humboldt-Universität zu Berlin, 12489 Berlin, Germany}
\affiliation[c]{Department of Physics, University of Florida, PO Box 118440, Gainesville, FL 32611-8440, USA}

\emailAdd{marek.kowalski@desy.de}
\emailAdd{markus.ackermann@desy.de}
\emailAdd{imrebartos@ufl.edu}

\date{January 2025}
\abstract{The detection of high-energy astrophysical neutrinos by IceCube has opened new windows for neutrino astronomy, but their sources remains largely unresolved. We study a methodology to address this -- deep-stacking -- that exploits correlations between observed neutrinos and comprehensive catalogs of potential source populations, including faint, high-redshift sources. By stacking signals from numerous weak sources and optimizing source weighting, significant gains in sensitivity can be achieved, particularly in the low-background regime where individual high-energy neutrinos dominate. We provide a semi-analytic framework to estimate sensitivity improvements for populations of sources under various background scenarios and redshift evolutions. Our analysis demonstrates that deep-stacking can increase detection sensitivity by a factor of 3–5, enabling detailed population studies. Furthermore, we discuss the potential to resolve the diffuse neutrino flux and investigate the redshift evolution of source populations. This approach offers a direct path toward identifying the primary sites of cosmic-ray acceleration and the mechanisms responsible for high-energy neutrino production.}

\begin{document}
%4\linenumbers

\maketitle

\section{Introduction}

%{\bf The classical \lnls{} distribution} can be understood from simple Euclidean geometry. For a source of constant luminosity $L$, the Flux $S\propto L/R^2$, where $R$ is the distance to the source. Accordingly, if a survey has a flux limit$S$, it will observe sources to a distance $R \propto S^{-1/2}$. The number of sources above a certain flux will be proportional to the cube of the survey radius, $N(>S)\propto S^{-3/2}$. The corresponding \lnls{} distribution therefore has a slope of -3/2.

%For gravitational waves, the detection of sources is based on the amplitude and hence scales as $R \propto S^{-1}$. The resulting number of sources grows correspondingly quick with sensitivity, $N(>S)\propto S^{-3}$. The corresponding \lnls{} distribution has a slope of -3.

Over the last decade, the flux of high-energy neutrinos detected by IceCube \cite{IceCube:2013low,IceCube:2014stg} has been thoroughly confirmed, with the spectrum measured within an energy range between 30\,TeV and 10\,PeV. Less clear is the nature of the sources producing the spectrum. Only a few sources have been identified so far: in 2017, coincident observations of neutrinos and gamma-rays from the Blazar TXS~0506+056 provided the first evidence for an extra-galactic neutrino source \cite{IceCube:2018dnn,IceCube:2018cha}. In 2022, IceCube provided evidence for high-energy neutrinos produced in  NGC 1068, one of the closest and brightest Active Galactic Nuclei (AGN; \cite{IceCube:2022der}). Other tentative observations include more correlations with Blazars and AGNs, as well as TDEs. 
 
Despite this progress, the current observations produce only an incomplete picture, as well as evidence that the nature of the neutrino sky is more complex (see e.g. \cite{Bartos:2021tok}). For instance, multiple independent analyses indicate that only a fraction of the observed neutrino flux originates from sources similar to those already identified by IceCube. Accordingly, it remains essential to identify the sources of the observed cosmic neutrinos. In fact, the discoveries of recent years raise a number of new, pressing questions that need to be answered, e.g., what are the conditions for cosmic ray (CR) acceleration and what are the mechanisms for neutrino emission. 

%One wonders what it takes to ultimately obtain a convincing answer. 
While neutrino astronomy offers a direct view on hadronic accelerators, it is so far unclear how to connect the neutrino and cosmic-ray observations beyond a qualitative picture. In particular, the connection to cosmic rays needs to be obtained from the modeling of the sources, which requires detailed multi-wavelength observations, as well as assumptions on the sources. Moreover, while neutrinos  travel on straight paths, detecting the sources comes with its own set of challenges, mainly due to the low rate of high-energy neutrino detections. Accordingly, a range of new observatories are being planned or are under construction \cite{KM3Net:2016zxf, P-ONE:2020ljt} and are expected to advance the field of neutrino astronomy, some with instrumented volumes an order of magnitude larger than IceCube  \cite{IceCube-Gen2:2020qha, Ye:2022vbk, Huang:2023mzt}.

Another route to gain sensitivity, which can be realized immediately, is to exploit the full information associated with the observed neutrinos in combination with detailed knowledge about prospective sources.
In this paper we explore the principle power of correlation analysis, allowing  inference of information concerning the sources of high-energy neutrinos. %the neutrino sky map, combined with the information about prospective sources,  as provided in detailed astronomical catalogs. 

As the signal-to-noise from prospective sources is figuratively speaking stacked on top of each other, the method is sometimes also referred to as {\it stacking analysis}. Initial larger stacking analyses have been performed in the past, including Blazars (see e.g.\ \cite{IceCube:2023htm}) or AGNs (e.g.\ \cite{IceCube:2021pgw}).
Most analyses performed so far have been somewhat limited in source catalog size and depth, as well as analyses methodology. 

In this paper, we discuss these shortcomings, and provide sensitivity estimates for optimally performed stacking analyses. We pay particular attention to the signal accumulated by stacking faint, high-redshift sources. Accordingly, we call the method laid out here {\it deep-stacking}. In a companion paper (PII) we will discuss the practical implementation of such a stacking analysis. %Our discussion will general encompass both bright as well as faint, sub-threshold neutrino sources, where the latter by themselves might not be significant. 

%Our method is applicable to both steady and variable source, such as Starburst galaxies or AGNs, or transients, such as Supernovae or Tidal Disruption Events (TDEs).  

%A special cass are sources  

Neutrino astronomy performed through deep-stacking can have a peculiar feature worth exploiting, as explained in the following using a variant of Olbers' paradox. Even faint sources can produce single high-energy cosmic neutrino events due to occasional Poisson fluctuations--an extreme form of Eddington bias \cite{nora_eddington}. In some cases, such events even allow for mildly significant associations with their source \cite{3sigma}. The rate of such events from nearby sources and distant sources can be comparable, since for large distances the drop in flux is just about compensated by having many more sources at the corresponding distances. Specifically, in a Euclidean Universe the flux per source ($S$) scales with distance ($R$) as $S \propto R^{-2}$. In estimating the detection rate of single neutrinos produced from within a certain distance shell, the drop in flux is compensated by the number of prospective sources, $dN_{\rm events} \propto S R^2 dR$. As a result, in the limit that each event allows to identify its source,  $N_{\rm sources}(<R)=N_{\rm events} \propto R \propto S(R)^{-1/2}$. The corresponding \lnls{} distribution for fixed $R$, therefore, has a slope of -1/2. This is very different from the slope one finds in the case of electromagnetic observations (see e.g. \cite{Ryle1955}). For the electromagnetic case, the number of observed sources will be proportional to the cube of the radius $R_0$ at which the sources' flux is at the observable threshold, i.e., $N(>S)\propto R_{0}^{3}\propto S^{-3/2}$. The corresponding \lnls{} distribution, therefore, has a slope of -3/2. The difference is due the large Eddington bias that helps identify the faint neutrino sources.

%For gravitational waves, the detection of sources is based on the amplitude and hence scales as $R \propto S^{-1}$. The resulting number of sources grows correspondingly quick with sensitivity, $N(>S)\propto S^{-3}$. The corresponding \lnls{} distribution has a slope of -3.

The rather flat \lnls{} distribution expected for neutrino sources identified through single neutrino associations is only an illustration of a general feature of multi-messenger neutrino astronomy. As we will show, using large catalogs of prospective counterparts for proper, deep-stacking analyses, significant gains can be expected also for associations with lower energy neutrinos that do not result in individual associations. In case the association cannot be performed with sufficient significance to make meaningful interpretations  on an event-by-event basis, one instead evaluates the correlation with a population of sources to improve the significance. We will show that  the redshift range covered by such an analysis is essentially only limited by the ability to detect the counterparts (until source confusion kicks in).   

This paper will provide the basic statistical arguments allowing to  forecast both the redshift dependence and the total sensitivity of a deep-stacking correlation analysis. It is organized as follows. In Sec.\ \ref{sec:nu_pop} we provide the mathematical framework to compute the sensitivity. In Sec.\ \ref{sec:nu_pop} we discuss requirements for the catalogs of prospective sources, as well as the potential to search for sub-populations of neutrino sources. Finally, in Sec.\ \ref{sec:cr} we provide a scientific motivation, including how the results of a deep-stacking analyses could provide novel information about the connection between CRs and neutrinos.

%In a second companion paper, we discuss more practical aspects of a search. 

\section{Neutrino source population sensitivity}
\label{sec:nu_pop}
We study the sensitivity of a generic neutrino telescope to identify  individual sources, as well as a population of sources using a semi-analytic approach. The practical implementation of the method can be found in PII.

We assume a standard candle neutrino source at a distance $d_j$, for which the flux $\phi_j$ is responsible for an expected number, $\hat{n}_s $,
signal neutrino events in a detector\footnote{We will indicate average event rates,  $\hat{n}=\left< n \right> $, using the hat-nomenclature.}
on top of a background of $\hat{n}_b$ events. The simplification associated with a standard candle will be discussed further in Sec.\ \ref{sec:catalog}.

Two estimators are typically used to determine the expected significance of a single source, given $\hat{n}_s$ and $\hat{n}_b$ . In the limit of $\hat{n}_s/\hat{n}_b\ll 1$, or the high-background case, the Fisher information method can be used to find the expected significance:
\begin{equation}
    \sigma_j = \left(\left< \frac{\partial^2 \log L(n,\hat{n}_s,\hat{n}_b)}{\partial \hat{n}_s^2}\right>\right)^{-1/2}=\hat{n}_s/\sqrt{\hat{n}_b},
    \label{eq:fisher}
\end{equation} where $\log L(n,\hat{n}_s,\hat{n}_b)$ is the Poisson likelihood function, that is being averaged over all possible number of observed events, $n$   (see e.g.\ \cite{fisher}). The Fisher method builds on the Taylor expansion of the log-likelihood function.  
The second method relies on the Asimov dataset \cite{Cowan:2010js}. It is applicable for any $\hat{n}_s/\hat{n}_b$, i.e. both for the low- and high-background cases. For the Asimov method, the expected sensitivity is
\begin{equation}
    \sigma_j = \sqrt{-2\ln\left(\frac{L(n_s=0,\hat{n}_b)}{  L(\hat{n}_s,\hat{n}_b)}
    \right)}
    =\sqrt{2 \cdot \left[(\hat{n}_s + \hat{n}_b)\ln(1+\hat{n}_s/\hat{n}_b)-\hat{n}_s\right]},
\end{equation} 
which reproduces the Fisher result of Eq.\ \ref{eq:fisher} for $\hat{n}_s/\hat{n}_b\ll 1$. 

As a first step, in this section we derive an analytic estimate of the sensitivity achieved in a stacking analysis using the simpler Fisher method provided in Eq.\ \ref{eq:fisher}. In the following sections, we will generally estimate sensitivity using the Asimov method, noting that for the discussed low-background case it gives similar result to Fisher's method.

A standard candle source at   distance $d_s$ will be observed with $\sigma_s \sim \sigma_j (d_j/d_s)^2$ significance, where we assume constant background expectation for any source direction. 
Next we consider a population of known sources of density $\rho$. On average, the distance of the nearest object will then be $d_n=(4/3 \pi \rho)^{-1/3}$.
There are more sources at larger distances/redshifts and the combined analysis from  all sources of the populations can be obtained through integration over redshift. Thereby we will use the fact that the combined sensitivity to $N$ identical sources is $\sigma^2=N\sigma_s^2$. Per (logarithmic) redshift shell, the significance is then 

\begin{equation}
    d\sigma^2/d\log z = z d\sigma^2/dz= z\sigma_s(z)^2 dN/dz,
    \label{eq:large_number}
\end{equation} 

where $dN/dz=\rho(z)dV/dz$  corresponds to the number of sources within a redshift shell. For a meaningful measure of the differential sensitivity we introduce the redshift dependent sensitivity for a population of sources:
\begin{equation}
    \sigma_z := \left[z d\sigma^2 /dz\right]^{1/2}. 
\end{equation}
The scaling with distance/redshift for a Euclidian Universe for a population of sources, $\sigma_z \propto z^{-1/2}$,  is notably different from that of a single source $(\sigma \propto z^{-2})$, 
  since the reduction of flux is partially compensated  by having more sources. The total significance to a population can then be obtained through integration, $\sigma^2=\int_{d_n} (d \sigma^2 /dz) dz$, where the integral is bound from below by the distance of the nearest object and from above either by the redshift range of the catalog, or, in case of a large number of prospective sources, source confusion.

%Using the Asimov dataset, the significance estimate now also holds  for very faint sources. 
In what  follows, we use the more general Asimov data for the significance of individual sources and combine it with Eq.\ \ref{eq:large_number} to generalize to a population of similar sources. 
%This quantity will be used   to illustrate the redshift dependent contributions to the cumulative sensitivity. 

\begin{figure}
    \centering
 \includegraphics[width=0.95\textwidth]{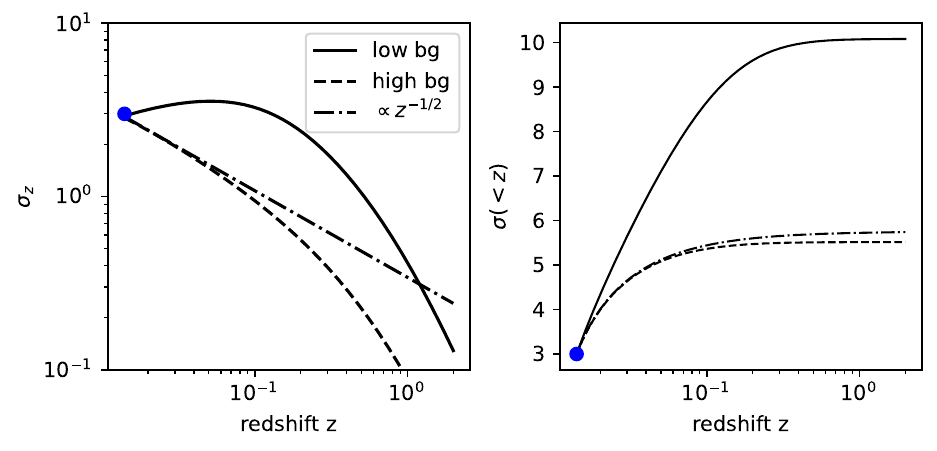}
 \caption{Left panel: differential sensitivity, $\sigma_z$, as a function of redshift  for a source population density,  $\rho =10^3~{\rm Gpc}^{-3}$, and normalized such that the expected significance of the nearest standard candle source is $3 \sigma$ (blue dot). All scenarios assume constant source number density and luminosity for all redshifts. The two scenarios probed correspond to the low and high-background scenario ($n_{\rm s}=1,n_{\rm bg=0.005}$; $n_{\rm s}=30,n_{\rm bg}=100$). In both case the Asimov statistics is used. The dash-dotted curve corresponds to the analytical estimate mentioned in the text.  Right panel: integral sensitivity as a function of the upper redshift cut-off.}
 
 %Besides the two different luminosity redshift evolutions, we show the expected significance of the standard candle point source at different redshift. 
\label{fig:sensi}
\end{figure}

In Fig.\ \ref{fig:sensi}, we show the differential and cumulative sensitivity as a function of redshift  for a source population density,  $\rho =10^3~{\rm Gpc}^3$ and no redshift evolution,  normalized such that the expected significance of the nearest standard candle source is $\sim3 \sigma$. 
Two scenarios are considered: First, a scenario with a large number of  background events ($n_{\rm sig}=30; n_{\rm bg}=100$) and second a scenario with low background, where single neutrinos from sources dominate the test statistics ($n_{\rm sig}=1; n_{\rm bg}=0.005$). We refer to these as the large-background and low-background scenario, respectively.  The cosmological volume evolution has been evaluated using cosmological parameters from the Planck mission \cite{planck18}. While in both case the Asimov statistics is used to compute the sensitivities, we  point out that for the large-$n$ scenario the Fisher method is also applicable. We also include the simple scaling in Fig.\ \ref{fig:sensi} using the analytical estimate for a Euclidian Universe obtained above above.  

It is obvious that even without redshift evolution, there is gain from properly accounting higher redshift sources. This is most striking for the singlet scenario, where the properties of the shallow \lnls{} distribution discussed in the introduction result in large gains due deep-stacking of sources.

%{\bf Simulation approach:}

%  Note so far we have assumed completness, as well as the complete classification of the corresponding source catalog. We will address that in Sec. \ref{sec:catalog}.

\section{Catalogs and proper weighting of sources}
\label{sec:catalog}

\begin{figure}
    \centering
 \includegraphics[width=0.95\textwidth]{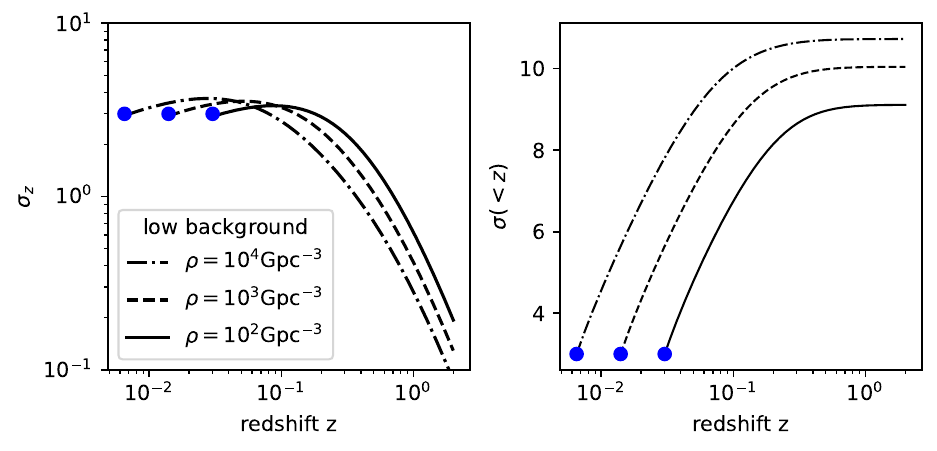}
 \caption{Left panel: differential sensitivity, $\sigma_z$, as a function of redshift  for three source population densities  $\rho =10^2,10^3,10^4~{\rm Gpc}^{-3}$, and normalized such that the expected significance of the nearest standard candle source is $3 \sigma$ (blue dot). All scenarios assume constant source number density and luminosity for all redshifts.}
 
 %Besides the two different luminosity redshift evolutions, we show the expected significance of the standard candle point source at different redshift. 
\label{fig:density}
\end{figure}

In this section, we discuss the issues associated with the catalogs required for the correlation analysis. One aspect is the completeness of the catalog. Parameterizing a catalog in terms of sky coverage, $\Omega$, and depth, $z_{\rm max}$, one finds that in most cases background levels are sufficiently high such that $\sigma_s \propto \Omega^{1/2}$. The dependence on depth is not expressed in such a closed form and instead shown in Fig.\ \ref{fig:sensi}.
We see in Fig.\ \ref{fig:sensi} that even without redshift evolution, the contributions from sources up to a redshift of $z \sim 0.3$ remain important. This is even more true in case of strong redshift evolution of source number density or luminosity, as will be discussed in Sec.\ \ref{sec:cr}. The local number density has also a mild impact on the redshift dependent sensitivity, as illustrated in Fig.\ \ref{fig:density} for three different cases. %For increasing number densities, the larger redshift sources contribute less. 

So far, we have ignored effects of source confusion, which will depend on the source number density and angular resolution.
This maximal distance can be estimated by requiring on  average one source per neutrino error contour: $d_{\rm max}\approx\left(4/3 \pi \rho \frac{\pi \psi_{1^\circ}^2}{4\cdot 10^4}\right)^{-1/3}\approx 23 d_n\psi_{1^\circ}^{-2/3}$, where $\psi_{1^\circ}$ corresponds to the angular resolution in degrees and $d_n$ the distance to the closest source.  For instance, the maximal distance for radio-loud AGNs with a local number density of $10^{3}$\,Gpc$^{-3}$ corresponds to $d_{\rm max}=1.4 {\rm Gpc}$, or $z_{\rm max} \approx d_{\rm max}c/H_0 \approx 0.3$, for an angular resolution of one degree. Larger source densities would need to be compensated by smaller angular resolution for the same distance reach. 
It is advisable to perform the correlation study with sufficiently deep catalogs.

%In case of TDEs this implies that in combination with LSST and ULTRASAT, we will be able to detect them beyond redshift $z\sim 0.5$. In case of Blazars, we can expect to identify coincidences even beyond $z\sim 1$. The essential aspect is that each coincidence observed provides an equal weighted $\frac{d\dot{N}}{dE_{\nu}}$ histogram entry, and hence allows to measure the integral on the right-hand-side of Eq. \ref{eq:time_integral}.

Next we discuss the optimal weighting of sources when stacking deeply. The optimal weight would thereby be proportional to the true, but unknown flux. Previously, we made the assumption of  standard candle neutrino sources, which implies a relative weight proportional to the event expectation, $\hat{n}_s \propto d_{j}^{-2}$. 
While this clearly is an unrealistic assumption, it's actually not the shape of the luminosity functions that matters that much for a qualitative statement. 
Instead, the key underlying assumption is that the relative flux between sources in the catalog is known, and only the overall normalization is unknown. For instance, this is implicit in the frequently-made assumption that the neutrino flux is proportional to the gamma-ray flux. 
This strong assumption then allows to perform an  inverse-variance weighting, resulting in an optimal estimator for the flux normalization. In any realistic scenario the relative flux is not precisely known and the sensitivity of a correlation analysis is expected to be lower, depending on the details of the luminosity function.
There will be an additional dispersion in luminosity, for which we will assume a log-normal luminosity function with width parameter, $\sigma_{\log f}$.

One can expect a minor effect on the sensitivity due to our a-priori ignorance of the luminosity, only if the variance from the Poisson process dominates the additional variance from the luminosity function. The condition is: $n_s<n_s^2 (e^{\sigma_{\log f}^2}-1)$, (the left hand side is the variance of the Poisson process, while the right hand side is that of a log-normal distribution).  
For $\sigma_{\log f}=1$, this corresponds to $n_s=0.6$. For brighter sources, the extra variance will reduce the statistical weight of the source, and hence the overall sensitivity.
Below this value, the extra variance due to the luminosity function is sub-dominant and hence does not change the sensitivity significantly.   When relying on single, high-energy neutrinos to make source associations, the condition is essentially always met. This is because of the large Eddington bias  discussed already in the introduction. \cite{nora_eddington} discusses the expected average event rate from sources producing only single neutrino events.  For a BL Lac source population, 95\% of the sources producing single neutrinos will have an underlying expectation of $n_s<0.6$, and hence meet our variance criterion. If the source population density is larger, the expectation per source will drop even further \cite{nora_eddington}. 

\begin{figure}
    \centering
 \includegraphics[width=0.95\textwidth]{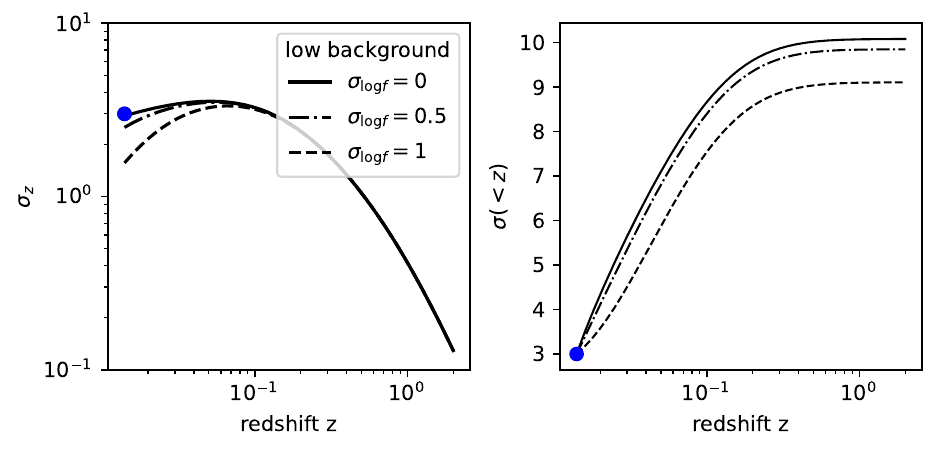}
 \includegraphics[width=0.95\textwidth]{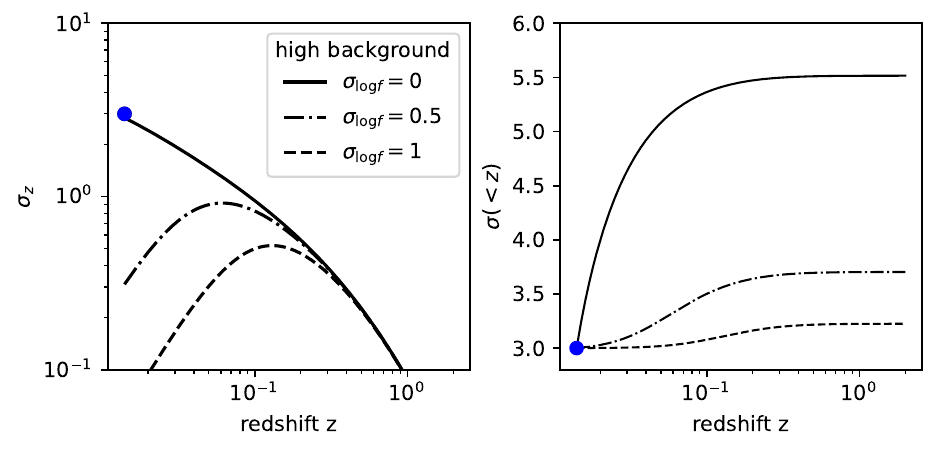}
 \caption{Left panel: differential sensitivity, $\sigma_z$, as a function of redshift  for a source population density (at z=0),  $\rho =10^3~{\rm Gpc}^3$, and normalized such that the expected significance of the nearest standard candle source ($\sigma_{\log f}=0$) is $\sim3 \sigma$ (blue dot). Also shown are curves with luminosity variance of  $\sigma_{\log f}=0.5,1$.  The signal and background represent the two cases shown in Fig.\ \ref{fig:sensi} (top: $n_{\rm s}=1,n_{\rm bg=0.005}$, bottom:  $n_{\rm s}=30,n_{\rm bg}=100$  ). Right panel: integral sensitivity as a function of the upper redshift cut-off. Note, the nearest source has been assumed to contribute at $3 \sigma$ independently of the luminosity variance.}
\label{fig:sensi_lumi}
\end{figure}

We illustrate the impact of the luminosity variance by recomputing the significance for a single source, $\sigma_s$ using an effective number of signal events, $n_s^\prime$, which is determined such that the total variance expected from the source  $V(n_s)=n_s+n_s^2 (e^{\sigma_{\log f}^2}-1)$ matches the Poisson variance of $V(n_s^{\prime}) =n_s^{\prime}$ in relative terms, $n_s^{\prime} = n_s^2/V(n_s)$.  The impact is shown in Fig.\ \ref{fig:sensi_lumi} for the low-background and  high-background scenarios introduced before.  

For the high-background scenario, the adjusted weighting that includes the luminosity variance leads to a  strong degradation of the significance at low redshifts, significantly reducing the total significance. 
As expected, the sensitivity in the low-background scenario  is only marginally affected by additional luminosity variance.

In conclusion, while the statistically optimal variance weighting appears to significantly affect low redshifts in the high-statistics case, assuming optimal weighting when working with single, high-energy neutrinos  is legitimate.   

\section{Identifying the population of neutrino sources}
So far we have assumed that a uniform catalog of  objects is provided, and that this is being tested for prospective neutrino emission. However, given that the catalog is curated based on only electro-magnetic information, without relying on a model for neutrino production, it can be expected that only a fraction of objects in the catalog will be strong neutrino emitters. Identifying this sub-population  is valuable for two reasons. First, scientifically, by identifying the production sights of high-energy neutrinos we gain a sharper view on the conditions required for neutrino production. This allows to better model the sources, as well as extrapolate to the neutrino emission from the full population, e.g.\ its contribution to the diffuse flux. Furthermore, we can also connect to cosmic ray production, as further discussed in Sec.\ \ref{sec:cr}. 

The second benefit is of statistical nature. For a fixed total number of neutrinos emitted by a population, the average neutrino flux per source will grow for a smaller sub-population. Accordingly, the sensitivity to the population will grow inversely proportional to the square-root of the density: $\sigma\propto \rho^{-1/2}$.  

But how should one identify the neutrino-emitting sub-population without relying on detailed model assumptions?
We assume the availability of estimates of the physical parameters for each object in a catalog. In the example case of AGN, this could be the SMBH mass, the bolometric luminosity or the obscuration. Armed with these parameters, one can perform the search for the best fitting sub-population as part of the neutrino-catalog cross-correlation analysis. For instance, this could imply scanning for the appropriate boundaries in the population parameter space. %the possibility to characterize the sub-population of neutrino emitters. 
For the sake of the argument we will  assume sharp boundaries, while in any realistic scenario they will be fuzzy to a certain level. We address this further below.

The shape of the volume in the parameter space will be a matter of choice, it could be a rectangular or elliptical volume that is one sided (e.g.\ SMBH mass above a certain value)  or two sided, hence adding between $n$ and $2n$ degrees of freedom for $n$ physical parameters that are evaluated. 

We evaluate the statistical penalty due to extra trial factors from scanning for unknown parameter boundaries vs the gain in signal-to-noise due to a reduced source density.  The problem at hand is that of a nested hypothesis test, where the null hypothesis corresponds to the parent population being the source of neutrinos, while the nested hypothesis corresponds to the sub-population defined through the extra parameters  being responsible for the neutrino flux.  Accordingly, we can use Wilks' theorem, which states that under the null hypothesis, the likelihood ratio will asymptotically follow the $\chi^2(n)$-distribution, where $n$ is the number of additional degrees-of-freedom describing the nested hypothesis. 

\begin{figure}

\begin{minipage}[b]{.49\textwidth}
    \includegraphics[width=0.99\textwidth]{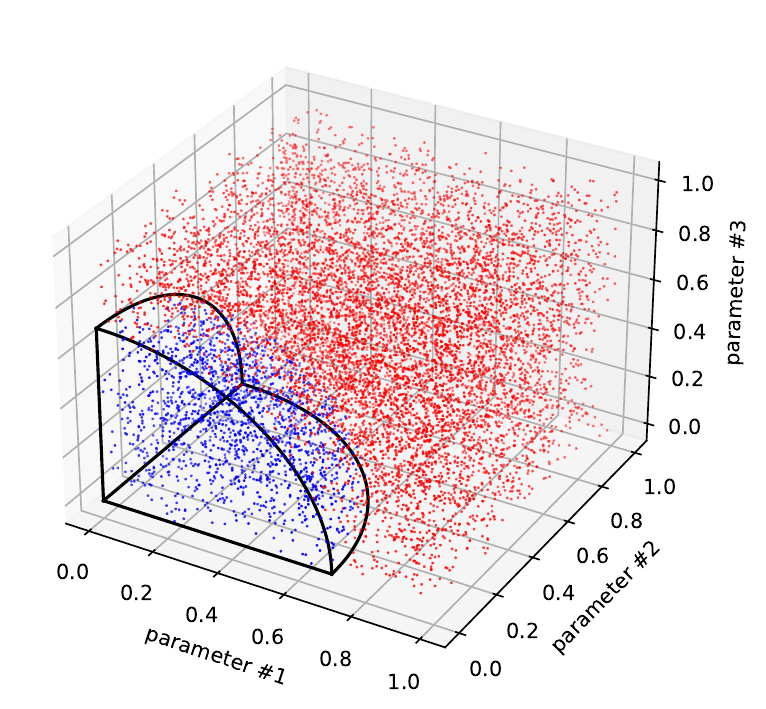}
\end{minipage}
\begin{minipage}[b]{.49\textwidth}\hfill
 \includegraphics[width=0.99\textwidth]{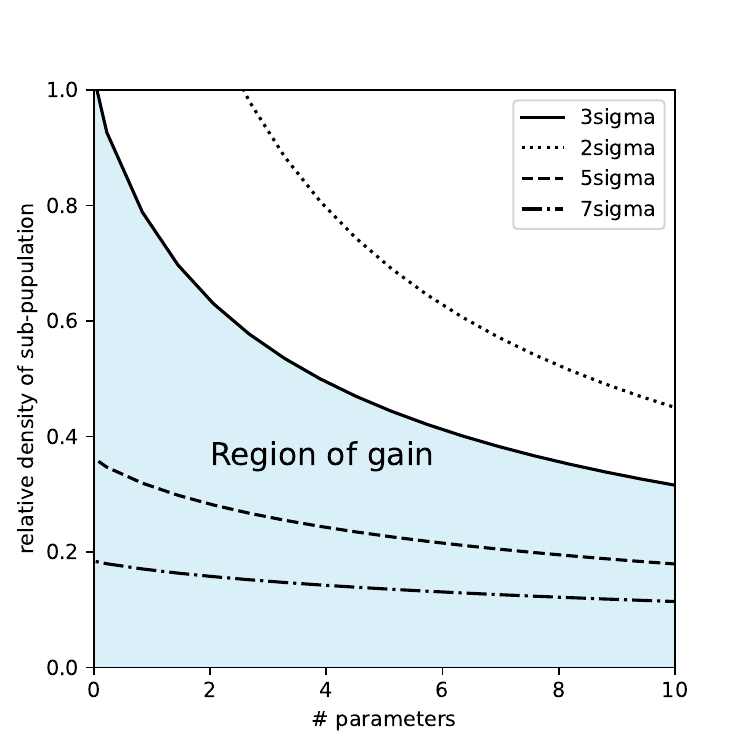}
\end{minipage}

 \caption{Left: Mock objects arranged according their  parameters. Blue / red dots indicate objects belonging to the signal / background population. Right: Significance as a function of number of parameters constraining a sub-population, as well as the relative density of the sub-population responsible for the signal.  }
\label{fig:wilkes}
\end{figure}

We evaluate the penalty due to the extra degrees of freedom vs. the prospective gain due to a reduced relative density in Fig.\ \ref{fig:wilkes}. There we have assumed that the parent population is already seen at $3\sigma$, hence a first hint has been observed already.

Whether there is a gain or not depends on the size of the subpopulation relative to its parent population. 
We find that in case the population of sources responsible for the neutrino signal is meaningfully smaller ( e.g.\ $<20\%$) relative to the parent population, the statistical gain overcompensates the penalty due to the extra  degrees of freedom (e.g.$\ n_{\rm d.o.f}\sim 10$). For smaller sub-populations, the gains could be significant, more than doubling the significance with which a population can be observed. 

The presence of observational uncertainties in the determination of the physical parameters will result in boundaries that are not sharp, and lead to sample contamination. 
We address the issue in the following in a general way. %that the information available for each source in a catalog, might not be sufficient for complete classification in a general way. 
Classification errors can be of type I (false-positives, contamination) or type II (false-negative, incompleteness). In the first case one would infer that the parent neutrino population is larger (by a factor $f_{\rm I}>1$), while in the second case, the catalog will only contain a fraction ($f_{\rm II}<1$) of neutrino sources. The classification errors will lead to a reduction in signal by $f_{\rm II}$, while the background will scale approximately as $f_{\rm I} f_{\rm II}$. Accordingly, the total signal-to-noise will scale as $(f_{\rm II}/f_{\rm I})^{1/2}$<1, a factor that is ideally close to one and which should be part of the consideration in any catalog selection/curation. 

If the classification errors can be controlled, the significant statistical gains should allow to identify a sub-population, where it exists.

\section{Applications}
\label{sec:cr}
So far we have outlined how to gain sensitivity to a population of sources, and how these gains consist of not just an overall increase in sensitivity, but also the ability to accumulate information on faint, sub-threshold sources. In the following, we outline three applications of the deep-stacking technique. 

\begin{figure}
    \centering
 \includegraphics[width=0.95\textwidth]{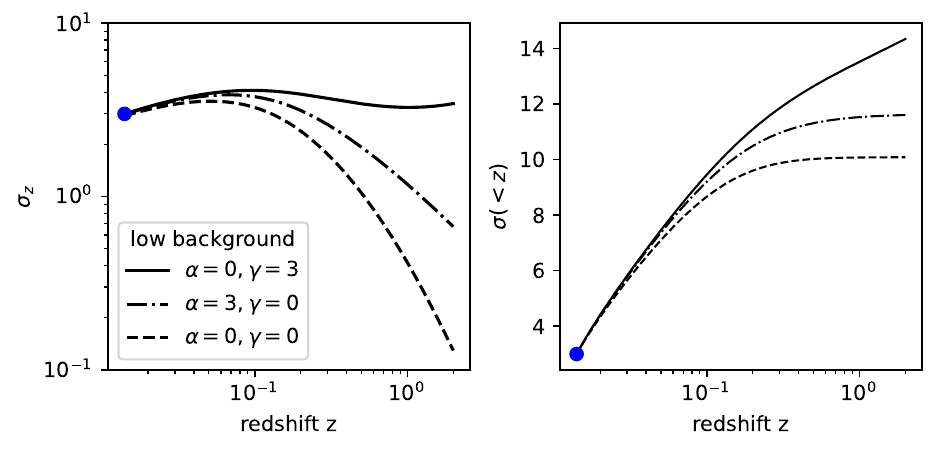}
 \includegraphics[width=0.95\textwidth]{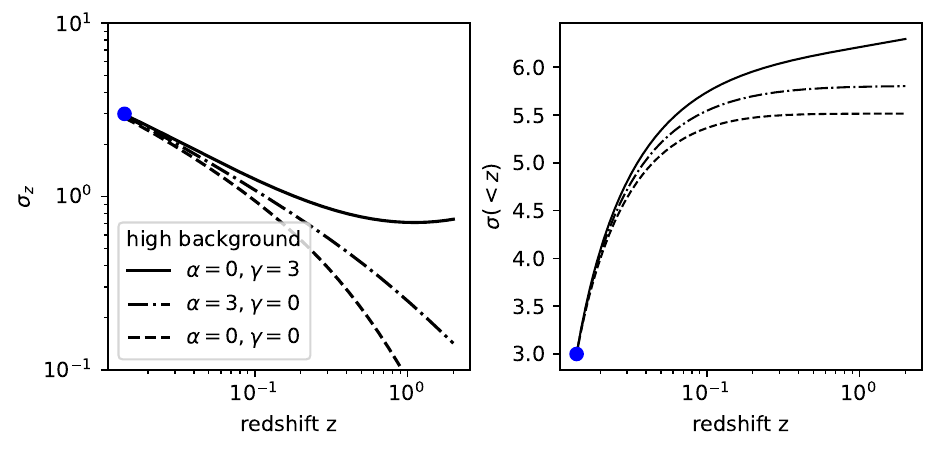}
 \caption{Left panel: differential sensitivity, $\sigma_z$, as a function of redshift  for a source population density (at z=0),  $\rho =10^3~{\rm Gpc}^3$, and normalized such that the expected significance of the nearest standard candle source is $\sim3 \sigma$ (blue dot). Redshift evolution of  luminosity $\propto (1+z)^\gamma$ and   density  $\propto (1+z)^\alpha$ is assumed. The total diffuse flux is the same for both cases with evolution. The signal and background represent the two cases shown in Fig.\ \ref{fig:sensi} (top: $n_{\rm s}=1,n_{\rm bg=0.005}$, bottom:  $n_{\rm s}=30,n_{\rm bg}=100$  ). Right panel: integral sensitivity as a function of the upper redshift cut-off.}
\label{fig:sensi_evolv}
\end{figure}

{\it 1) Redshift evolution of source density and source luminosity:}
Identifying a source population goes hand in hand with learning about the redshift evolution of both density and luminosity of its' sources.  The techniques presented here will provide sensitivity in two ways. First, by identifying the  sub-population responsible for the neutrino emission, the decisive properties of the sources will be identified. Having a characterization of the sources will allow us to obtain the redshift evolution from the catalog information, only.% (this assumes that the catalog information reaches sufficiently deep). 

Second, the redshift evolution can be measured directly (see Fig.\ \ref{fig:sensi_evolv}). We had introduced redshift evolution parameters and have shown that there is sensitivity to those, at least in the case of strong redshift evolution. In terms of signal-to-noise, strong redshift evolution of the luminosity (such as expected for certain AGN accretion phenomena) will result in a larger significance compared to  strong density evolution (such as expected for Supernovae). In case the redshift evolution is weak or even negative (such as expected for TDEs), one expects little signal  to be contributed at high-redshifts, and hence this scenario can be identified by an absence of a correlation signal at high redshifts.

{\it 2) Resolving the diffuse neutrino flux:} Once a population of sources is identified an immediate follow-up question concerns its relative contribution to the total diffuse flux. Are there undetected neutrino source populations remaining, or are the contributions from the resolved populations sufficient to explain the diffuse flux? The answer to this question follows directly from the best fit cumulative spectrum obtained from the deep-stacking analysis. Parameters such as those related to the redshift dependent source density or luminosity evolution can thereby be marginalized, since only the cumulative flux from all sources in a catalog matters.      

{\it 3) The sources of high-energy cosmic rays:}
We address the question of whether the extragalactic neutrino sources could also be the sources of high-energy cosmic rays at the highest energies. Such arguments for populations of sources where already used in the broadest terms to establish an upper bound on the cosmic neutrino flux derived from the observed cosmic ray flux, based on the limited efficiency for converting cosmic rays to neutrinos \cite{Waxman:1998yy}. Building on the above applications 1) and 2), one can  study the relationship observationally.   While for individual sources, such as NGC\,1068 or TXS\,0506+056, detailed modeling enable us to infer information about the underlying cosmic ray spectrum, these are just individual sources observed during a short epoch of the universe. This is starkly different from cosmic rays observed on Earth.
The cooling time of extragalactic cosmic rays up to energies of $10^{18}$ eV corresponds to the Hubble time. Accordingly, extragalactic cosmic rays observed on Earth today will  have been produced  through cosmic times.

One can relate the flux of cosmic rays to the production rate of cosmic rays and neutrinos, integrated over cosmic times:

\begin{equation}
I_{\rm CR}=\int_{t_0} \frac{d\dot{N}}{dE_{\rm CR}} dt \approx \epsilon c \int_{z=0}  E(z)^{-1}\frac{d\dot{N}}{dE_{\nu}}dz,
\label{eq:time_integral}
\end{equation}

Here, $\epsilon$ is the production efficiency for neutrinos from cosmic rays and $E(z)$ is just a function of cosmological parameters, \begin{equation}
    E(z) = (1+z) H_0 \left[ \Omega_M (1+z)^3+ \Omega_\Lambda \right]^{1/2}.
\end{equation}  

Note  that $E_{\nu} = 0.05 E_{\rm CR}$, thus the PeV energies we currently probe with IceCube correspond to $\sim 10$ PeV CR energies, close to the transition energy of galactic and extra-galactic cosmic rays.
What we have shown in Sec. \ref{sec:nu_pop} is that for selected sources, one can measure the redshift dependent flux $\phi = \frac{d\dot{N}}{dE_{\nu}}$, assuming a constant spectrum. 
With a sufficiently well characterized population of neutrino sources, we should be able to establish the production sites and environments, and hence determine the efficiency, $\epsilon$, through modeling. We can also translate this to the corresponding CR production history, providing us with a direct path to the answer, whether the neutrino sources observed can explain the observed flux of cosmic rays.  

%We consider only an idealized relation: 
%\begin{equation}
%    E_{\nu}^2 \frac{d\dot{N}}{dE_{\nu}} = \epsilon E_{\rm CR}^2 \frac{d\dot{N}}{dE_{\rm CR}}.
%\end{equation}

%\section{Conclusion}
%...
%This deep reach into cosmic history provides significant opportunities for the study of extreme accretion and particle acceleration processes. We have layed out a blueprint how to connect the redshift resolved neutrino flux to the diffuse neutrino and CR flux observed today. 

%By precisely measuring the extragalactic CR flux and composition above $10^{16}$~eV, along with studying the sources of cosmic neutrinos using mulimessenger techniques across different redshifts, we measure both sides of the equation, and hence can comprehensively settle the question of the origin of extragalactic CRs. 

\section{Conclusions}

We have studied the achievable gains in the search for high-energy neutrino sources using the  cross-correlation signal between high-energy neutrinos and comprehensive catalogs of astronomical objects. Particular attention is given to recovering the information from the many faint, high-redshift sources that are often emitted from correlation studies. Accordingly, we refer to such an analysis as deep-stacking. Conventional searches for individual point sources or stacking of multiple  sources use a log-likelihood approach that also includes a neutrino energy estimator, which, for high energies, allows us to preferentially select cosmic neutrinos from the large background of atmospheric neutrinos. We discuss the low-energy and high-energy regime by providing two characteristic scenarios, one with high background and the other with low background. (In reality, one can expect to have a mixture, with one dominating the sensitivity).
Key findings of our study are:
\begin{enumerate}
    \item Neutrino astronomy in the low-background regime significantly profits from the deep-stacking of sources, This is due to the large Eddingtion bias for these sources, which results in a peculiarly flat \lnls{} distribution. The high-background case, which relies on neutrinos at lower energies, which can be identified above a large background through multiplets, is less promising for the deep-stacking of sources.   
    
    \item For a population of neutrino sources, unaccounted variance in neutrino luminosity will lead to sub-optimal weighting in the stacking analysis.  In the high-background regime, variance in neutrino luminosity will reduce the sensitivity rather significantly. In contrast, the low-background regime is almost immune to this effect.
    
    \item The search for correlation benefits from comprehensive and complete catalogs of astronomical sources. We find that the distance range over which most of the sensitivity gain occurs is out to redshift $z \sim 0.3$. This is particularly advantageous for deep-stacking, as reliable and complete catalogs are readily available within this interval.
    
    \item In the ideal case of a complete catalog the gains in sensitivity would amount to a factor of 3-5 in sensitivity,  depending on if there is redshift evolution or not and on the hardness of the neutrino spectrum. 
    
    \item A comprehensive study should enable the systematic determination of the neutrino source population, including possible sub-population of sources as well as the redshift evolution of both density and luminosity. The applications range from resolving the source populations responsible for the the diffuse high-energy neutrino flux, to the study of the history of cosmic ray production.
    
\end{enumerate}

%In summary, we find that despite the challenges to identify individual high-energy neutrino sources, deep-stacking offers a route towards studying source populations. It requires complete and comprehensive catalogs of astronomical objects. The  sensitivity mainly derives from the high-energy tail of the neutrino spectrum, which is characterized by its low backgrounds. Prospective rewards of a properly performed  deep-stacking search include a high-significance detection of a population of high-energy neutrino sources, identification of prospective sub-populations, and its characterization of redshift evolution. 

In summary, we find that despite the inherent challenges in pinpointing individual high-energy neutrino sources, deep-stacking provides a pathway for studying source populations with improved sensitivity. By leveraging complete and detailed catalogs of astronomical objects, the technique maximizes the information derived from faint, distant sources that typically remain unconsidered in conventional analyses. This approach primarily benefits from the high-energy tail of the neutrino spectrum, where backgrounds are lower. Beyond enhancing sensitivity, deep-stacking enables new scientific opportunities, including identifying sub-populations of sources, characterizing their redshift evolution, and exploring the connection between high-energy neutrinos and cosmic-ray production. Ultimately, this method offers a promising avenue toward a comprehensive understanding of high-energy astrophysical processes and the nature of the diffuse neutrino flux.

\acknowledgments
We are grateful to Jakob van Santen for providing helpful comments. IB is grateful for the support of the National Science Foundation under grant No. PHY-2309024.

\bibliographystyle{JHEP}
\bibliography{my-bib.bib}

\end{document}